\documentclass[11pt,english]{article}
\usepackage[T1]{fontenc}
\usepackage[latin9]{inputenc}
\usepackage{geometry}
\usepackage{amsmath}
\usepackage{amssymb}

\newcommand{\Tr}{\mathrm{Tr}}
\newcommand{\diag}{\mathrm{diag}}
\newcommand{\CSA}{\mathrm{CSA}}

\makeatletter
\numberwithin{equation}{section}

\date{} 
\usepackage[T1]{fontenc} 
\usepackage{pslatex} 

\makeatother

\usepackage{babel}
\begin{document}
\setcounter{page}{0}

$\,\,$\vspace{1cm}

\begin{center}
\textbf{\LARGE{}Continuous families of $\mathbb Z_2$ monopoles in SU(4) Yang-Mills-Higgs theory }
\par\end{center}{\LARGE \par}

\vspace{1cm}

\begin{center}
\textbf{Paulo J. Liebgott}\footnote{paulo.liebgott@ufsc.br}\textbf{
and Eduardo E. Quadros}\footnote{eduardoequadros@gmail.com}
\par\end{center}

\begin{center}{\em Departamento de F\'\i sica,\\ 

Universidade Federal de Santa Catarina (UFSC),\\

Campus Universit\'ario, Trindade,\\

88040-900, Florian\'opols, Brazil.  \\ }

\end{center}

\vspace{0.3cm}
\begin{abstract}
We consider an SU(4) Yang-Mills-Higgs theory spontaneously broken to $SO(4)$ by a scalar field in the $n\times n$ representation and the unbroken algebra invariant under Cartan automorphism. We obtain that $\mathbb Z_2$ monopoles in this theory belong to different classes of $su(2)$ embeddings associated to continuous families of up to four parameters. This result generalizes the discrete families of $\mathbb Z_2$ monopoles previously known as well as it shows new classes of embeddings.

\vfill 

\thispagestyle{empty}
\end{abstract}

\newpage

\section{Introduction}

The color confinement problem in QCD \cite{Greensite:2011zz} is one of the most challenging problems in theoretical physics and according to the ideas of 't Hooft and Mandelstam, monopoles may be relevant to its solution \cite{tHooft1975,Mandelstam:1974pi}. They conjectured that electric charged particles would be confined by a dual Meissner effect in the same way magnetic charges would be confined in a superconductor \cite{tHooft:1982ylj}. However, one of the main ingredients of this conjecture, the electromagnetic duality, is an unsolved problem by itself \cite{OLIVE199688}. Although there are some well known results about it \cite{GNO1977,MO1977,VW1994,SW1994_1,SW1994_2}, there is no proof of an exact electromagnetic duality in non-Abelian theories. The concept of magnetic monopoles may be useful for both problems and therefore a good understanding of such objects is required.
 
Monopoles of topological nature are predicted by spontaneously broken non-Abelian gauge theories whose vacuum manifold has non-trivial second homotopy group \cite{GO1978,manton2004topological,WEINBERG200765}. The elements of this group classify topological solutions into equivalence classes and it is impossible to continuously deform one solution into another if they belong to different classes. The simplest and best known example this type of monopole is the 't Hooft-Polyakov monopole \cite{tHooft1974,Polyakov1974}. It appears in $SO(3)$ Yang-Mills-Higgs theory spontaneously broken to $SO(2)$ by a scalar field in the adjoint representation and its topological classes belong to $\mathbb Z$, the group of integers. These monopoles are known as $\mathbb Z$ monopoles.

When the the second homotopy group of the vacuum manifold is the cyclic group $\mathbb Z_n$, the theory may have the so-called $\mathbb Z_n$ monopoles. Normally, they appear when the scalar field is in a representation other than the adjoint. There are some known results about $\mathbb Z_n$ monopoles \cite{GNO1977,Weinberg1983,BL1988} but these objects are less studied than the $\mathbb Z$ monopole. In particular it is known how to obtain asymptotic field configurations for $\mathbb Z_n$ monopoles. Just like for 't Hooft-Polyakov monopoles, the field configurations of the $\mathbb Z_n$ monopole live in an appropriated $su(2)$ subalgebra. For 't Hooft-Polyakov monopoles this task is trivial since there is only one possible $su(2)$ embedding. For larger algebras there are more possibilities though. The finding of these embeddings is therefore a crucial part of the work of explicitly constructing $\mathbb Z_n$ monopoles solutions.
 
In \cite{KL2010,KL2013} the authors found, in particular, that $SU(2m)$ Yang-Mills-Higgs broken to $SO(2m)$ with the unbroken algebra invariant under Cartan automorphism have finite families of BPS $\mathbb Z_2$ monopoles solutions. They obtained a total of $(2^{m+1}-2)$ monopoles. In the present work we devise a method to generalize those results and obtain larger families of $\mathbb Z_2$ monopoles. We show that those monopoles belong to continuous families and we find all possible families for the case $m=2$.

This work is organized as follows: in section \ref{sec:review} we briefly review some known results for topological monopoles and set up some mathematical conventions. In section \ref{sec:algebras} we present a general method of obtaining classes of continuous families of appropriate $su(2)$ subalgebras from which we can construct asymptotic monopole solutions. The possible embeddings for $SU(4)\rightarrow SO(4)$ are explicitly obtained in section \ref{sec:classes}. In section \ref{sec:conclusions} we present our concluding remarks.


\section{Theoretical framework}\label{sec:review}

In this sections we shall recall some general results about topological monopoles which are going to be useful throughout this paper. We also set up notation and mathematical conventions. Let us start by considering Yang-Mills theories with Lagrangian
$$
\mathcal{L}=-\frac{1}{4}G_{\mu\nu}^{a}G^{a\mu\nu}+\frac{1}{2}(D_{\mu}\phi)_{a}^{\ast}(D^{\mu}\phi)_{a}+V(\phi),
$$
where $\phi$, is a scalar fields in some representation of the gauge group $G$. We assume these theories are spontaneously broken to the unbroken gauge group $G_0$ and admit topological magnetic monopoles. These monopoles must satisfy a non-Abelian quantization condition \cite{EW1976}
\begin{equation}
\exp(ieg)= I,
\label{eq:quantization_condition}
\end{equation}
where 
\begin{equation}
g=\frac{4\pi}{e}T_3=\frac{2\pi}{e}\beta\cdot h,\label{eq:magnetic_charge}
\end{equation}
is the general form of the non-Abelian magnetic charge and $h=(h_1,h_2,\ldots,h_r)$ denotes the Cartan subalgebra generators of the unbroken algebra \cite{GNO1977}.

If $G=SO(3)$ is spontaneously broken to a subgroup $G_0=SO(2)$ by a scalar field in the adjoint of $SO(3)$ then the vacuum manifold has a nontrivial second homotopy group, namely $\pi_2(SO(3)/SO(2))\cong \mathbb Z$, and the theory presents the so-called 't Hooft-Polyakov monopoles \cite{tHooft1974,Polyakov1974}. By following the clever construction due to Weinberg \textit{et al.} \cite{Weinberg1983} we can obtain the asymptotic fields for monopoles with winding number $q$ as
\begin{equation}
\begin{aligned}
\phi(\theta,\varphi) & =g(\theta,\varphi)\phi_0g(\theta,\varphi)^{-1},\\
W_{i}(\theta,\varphi) & =g(\theta,\varphi)W_i^{\mathrm{string}}g(\theta,\varphi)^{-1}-\frac{i}{e}\left(\partial_{i}g(\theta,\varphi)\right)g(\theta,\varphi)^{-1},\\
W_{0}(\theta,\varphi) & =0,
\end{aligned}
\label{eq:general asymptotic fields}
\end{equation}
where
$$
\begin{aligned}
\phi_0&=vT_3,\\
W_r^{\mathrm{string}}&=W_\theta^{\mathrm{string}}=0,\\
W_\varphi^{\mathrm{string}}&=-\frac{T^3}{er}\frac{(1-\cos\theta)}{\sin\theta},
\end{aligned}
$$
and
$$g(\theta,\varphi)=\exp(-iq\varphi T_3)\exp(-i\theta T_2)\exp(iq\varphi T_3).$$
Here, the generators $T_a$, $a=1,\, 2,\, 3$, form an appropriate $su(2)$ subalgebra such that $T_3$ is given by \eqref{eq:magnetic_charge} and the remaining two generators do not belong to the unbroken algebra. 

Due to Weinberg's construction we say there is an $su(2)$ subalgebra or embedding associated to each monopole. Then an important point to notice is that there is a continuous one-parameter family of possible $su(2)$ subalgebras used to construct the 't Hooft-Polyakov monopole in the above theory. By applying the global transformation generated by $\exp(i\chi T_3)$ we obtain the asymptotic solutions given by \eqref{eq:general asymptotic fields} with new generators
\begin{equation}\label{eq:tHP1}
T_1'=\cos\chi T_1+\sin\chi T_2,\quad T_2'=-\sin\chi T_1+\cos\chi T_2,\quad T_3'=T_3.
\end{equation}
This $U(1)$ symmetry is the only internal degree of freedom available in the 't Hooft-Polyakov solution and is one of the four zero modes of the unit charge 't Hooft-Polyakov monopole \cite{Weinberg1979}.

For other gauge groups, as well as other representations of the scalar field, there might be the so-called $\mathbb Z_n$ monopoles \cite{Weinberg1983,BL1988}. In this work we consider $SU(n)$ Yang-Mills-Higgs theories spontaneously broken by the vacuum configuration
\begin{equation}\label{eq:vacuum}
\phi_0=\frac{v}{\sqrt2}\sum_{k=1}^{n}|k,k\rangle,    
\end{equation}
which is in the $n\times n$ representation of the gauge group. These theories have the topological conditions allowing the existence of monopoles \cite{KL2010,KL2013}. In that case, the unbroken group is $SO(n)$ and the vacuum manifold is nontrivial, $\pi_2(SU(n)/SO(n))\cong\mathbb Z_2$, giving rise to $\mathbb Z_2$ monopoles. The quantization condition \eqref{eq:quantization_condition} is equivalent to the magnetic weight $\beta$ belonging to either one of the cosets $\Lambda_r(so(n)^\vee)$ or $\lambda_1^\vee+\Lambda_r(so(n)^\vee)$, where $\Lambda_r(so(n)^\vee)$ denotes the root lattice of the dual algebra $so(n)^\vee$ and $\lambda_i^\vee$ is a fundamental weight $so(n)^\vee$, and the solutions associated to the latter are in the nontrivial topological sector. We recall that $so(2m)^{\vee}=so(2m)$, $so(2m+1)^{\vee}=sp(2m)$ and the $2m$ weights of the defining representation of both algebras can be written terms of the unit Cartesian vectors $\pm e_a$, $a=1,\ldots,m$. Then the generators
$$
T_3=\pm \frac{e_a\cdot h}{2}=\pm\frac{(E_{\alpha_{2a-1}}-E_{-\alpha_{2a-1}})}{2i}\in so(n),
$$
where $E_{\alpha_{2a-1}}$ are $su(n)$ step operators associated to simple roots $\alpha_{2a-1}$, satisfy the non-Abelian quantization condition and there are $2m$ fundamental monopoles associated to the magnetic weights $\pm e_a$. Their asymptotic solutions are obtained as $su(2)$ embeddings in $su(n)$ in a similar way it is done for 't Hooft-Polyakov monopole. The known $su(2)$ subalgebras for the fundamental monopoles are given by
\begin{align*}
T_1^{\pm e_a}&=\pm\frac{1}{2}( E_{2a-1,2a-1}-E_{2a,2a})=\pm\frac{1}{2}e_{aa}\otimes\sigma_3,\\
T_2^{\pm e_a}&=\frac{1}{2}(E_{2a-1,2a}+E_{2a,2a-1})=\frac{1}{2}e_{aa}\otimes\sigma_1,\\
T_3^{\pm e_a}&=\pm\frac{1}{2i}( E_{2a-1,2a}-E_{2a,2a-1})=\pm\frac{1}{2}e_{aa}\otimes\sigma_2.
\end{align*}
Since we are going to work at the level of the algebra, it is convenient to adopt the defining representation. Therefore $(E_{ij})_{kl}=\delta_{ik}\delta_{jl}$. Moreover, $e_{ab}$ stands for the $m\times m$ version of $E_{ij}$ and $\sigma_i$, $i=1,2,3$ are the Pauli matrices. Through this paper, we assume indices $i,j,k,\ldots$ run from $1$ to $2m$ whereas $a,b,c,\ldots$ run from $1$ to $m$ unless something else is specified. There is no summation over repeated indices unless there is an explicit summation symbol.  

There are also known nonfundamental $\mathbb Z_2$ monopoles which are combinations of the previous ones, constructed from
$$T_1^{\pm\beta}=\sum_{a}n_aT_1^{\pm e_a},\quad T_2^{\pm\beta}=\sum_{a}n_aT_2^{\pm e_a},\quad T_3^{\pm\beta}=\sum_{a}n_aT_3^{\pm e_a},$$
where $n_a=0,1$. Notice that the above subalbegras are linear combinations of the subalgebras associated to the fundamental monopoles and they do not mix positive and negative $\pm e_a$. 

As particular case of the above example we consider $G=SU(4)$ spontaneously broken to $SO(4)$. This theory has four fundamental monopoles in one-to-one correspondence to the following $su(2)$ subalgebras
\begin{equation}\label{eq:KL1}
\left\lbrace
\begin{aligned}
T_1^{\pm e_1}&=\pm\frac{1}{2}( E_{11}- E_{22})=\pm\frac{1}{2}e_{11}\otimes\sigma_3,\\
T_2^{\pm e_1}&=\frac{1}{2}( E_{12}+ E_{21})=\frac{1}{2}e_{11}\otimes\sigma_1,\\
T_3^{\pm e_1}&=\pm\frac{1}{2i}( E_{12}- E_{21})=\pm\frac{1}{2}e_{11}\otimes\sigma_2,
\end{aligned}
\right.
\quad
\left\lbrace
\begin{aligned}
T_1^{\pm e_2}&=\pm\frac{1}{2}( E_{33}- E_{44})=\pm\frac{1}{2}e_{22}\otimes\sigma_3,\\
T_2^{\pm e_2}&=\frac{1}{2}( E_{34}+E_{43})=\frac{1}{2}e_{22}\otimes\sigma_1,\\
T_3^{\pm e_2}&=\pm\frac{1}{2i}( E_{34}- E_{43}))=\pm\frac{1}{2}e_{22}\otimes\sigma_2,
\end{aligned}
\right.
\end{equation}
and two nonfundamental monopoles in one-to-one correspondence to the $su(2)$ subalgebras below
\begin{equation}\label{eq:KL2}
\begin{aligned}
T_1^{\pm (e_1+e_2)}&=\pm\frac{1}{2}(E_{11}- E_{22}+ E_{33}- E_{44})=\pm\frac{1}{2}\sum_ae_{aa}\otimes\sigma_3,\\
T_2^{\pm (e_1+e_2)}&=\frac{1}{2}( E_{12}+ E_{21}+ E_{34}+ E_{43})=\frac{1}{2}\sum_ae_{aa}\otimes\sigma_1,\\
T_3^{\pm (e_1+e_2)}&=\pm\frac{1}{2i}( E_{12}- E_{21}+ E_{34}- E_{43})=\pm\frac{1}{2}\sum_ae_{aa}\otimes\sigma_2.
\end{aligned}
\end{equation}
The important question we could ask is whether those are the only $su(2)$ subalgebras possible, hence the only monopoles, or not. That is the question we are addressing to in this work.


\section{New $su(2)$ subalgebras}\label{sec:algebras}

In order to find new $su(2)$ embeddings it is easier to work with generators in the defining representation. We can do this because we are working at the level of the algebra. Moreover, since we are interested in the $SU(4)\rightarrow SO(4)$ case, we will focus on the even dimensional case $SU(2m)\rightarrow SO(2m)$. 

Let us begin by finding the general form of
$$T_3=\frac{\beta\cdot h}{2}\in \CSA(so(n)),$$
where $\CSA$ denotes Cartan subalgebra, satisfying the non-Abelian quantization condition. Since $T_3$ belongs to the unbroken algebra, it automatically annihilates the vacuum configuration \eqref{eq:vacuum}. Writing the magnetic weight weight as $\beta=\sum_a\beta_a e_a$, $a=1,\ 2,\ldots,\,  m$, equation \eqref{eq:quantization_condition} implies that the coefficients $\beta_a$ are integers. We also know that the generators of $\CSA(so(2m))$ can be given by $h_a=E_{2a-1,2a}-E_{2a,2a-1}$. Then it proves useful to factor $h_a$ as the tensor product $h_a=-i\diag(e_a)\otimes J,$
where $J$ is the $2\times 2$ symplectic form,
\begin{equation}\label{eq:J}
J=
\left(
\begin{array}{cc}
0&1\\
-1&0
\end{array}
\right)=i\sigma_2.
\end{equation}
Hence
\begin{equation}\label{eq:T3}
T_3=\frac{1}{2i}\diag(\beta)\otimes J=\frac{1}{2}\sum_{ab}\beta_a\delta_{ab}e_{ab}\otimes \sigma_2,\quad \beta=\sum_{a}\beta_ae_a ,\quad \beta_a\in\mathbb Z.
\end{equation}

The remaining generators of $su(2)$ can be obtained in the following way. We first eliminate $T_2$ from the system of equations $[T_i,T_j]=i\epsilon_{ijk}T_k$ to reach the second order system
\begin{subequations}\label{eq:system1}
\begin{align}
[T_3,[T_3,T_1]]&=T_1,\label{eq:system1a}\\
[[T_3,T_1],T_1]&=T_3.\label{eq:system1b}
\end{align}
\end{subequations}
Then solving the above equations for $T_1$ and $T_3$ we are able to obtain $T_2$ by 
\begin{equation}\label{eq:T2}
T_2=i[T_1,T_3].
\end{equation}
We have to bear in mind that $T_1$ and $T_2$ must not belong to the unbroken algebra though \cite{Weinberg1983}.

In order to solve the system \eqref{eq:system1} for $T_1$, it is convenient to write this generator as
\begin{equation}\label{eq:T1}
T_1=\frac{1}{2}\sum_{ab}e_{ab}\otimes x_{ab},
\end{equation}
where $x_{ab}$ are $2\times 2$ blocks to be determined. Notice that $x_{ab}^\dagger=x_{ba}$ and $\Tr (x_{aa})=0$ because $T_1$ is Hermitian and traceless. From \eqref{eq:T3}, \eqref{eq:system1a} and \eqref{eq:T1} we obtain
$$T_1=\frac{1}{8}\sum_{ab}e_{ab}\otimes(\beta_a^2x_{ab}+2\beta_a\beta_bJx_{ab}J+\beta_b^2x_{ab}).$$
We compare the above equation with \eqref{eq:T1} and since $e_{ab}$ are all linearly independent we get the following fundamental relation satisfied by blocks $x_{ab}$ and magnetic weight components
\begin{equation}\label{eq:beta}
(4-\beta_a^2-\beta_b^2)x_{ab}=2\beta_a\beta_bJx_{ab}J.
\end{equation}
For at least one pair $(a,b)$ we must have $\beta_a\beta_b\neq 0$, otherwise the magnetic weight would vanish identically. Now, notice that for any complex-valued $2\times 2$ matrix $x$ and $J$ given by \eqref{eq:J}, the solution of $JxJ=kx$ is either $k=\left\lbrace 1,-1\right\rbrace$ or $x=0$. Applying this result to \eqref{eq:beta} we obtain
\begin{equation}\label{eq:beta2}
4-\beta_a^2-\beta_b^2=2k_{ab}\beta_a\beta_b,
\end{equation}
for $\beta_a,\beta_b\neq 0$, $k_{ab}=k_{ba}\in\left\lbrace 1,-1\right\rbrace$ and nonvanishing block
\begin{equation}\label{eq:blocks}
x_{ab}=\left(
\begin{array}{cc}
u_{ab}&v_{ab}\\
k_{ab}v_{ab}&-k_{ab}u_{ab}
\end{array}
\right),
\end{equation}
with $u_{12},v_{12}\in\mathbb C$. Recalling that the magnetic weights coefficients must be integers, we can write \eqref{eq:beta2} as
\begin{equation}\label{eq:beta3}
l_{ab}\beta_a+l_{ba}\beta_b=2,\quad \beta_a,\beta_b\neq 0,\quad x_{ab}\neq 0,
\end{equation}
for signs $l_{ab}$ and $l_{ba}$ satisfying $k_{ab}=l_{ab}l_{ba}$. The case $a=b$ readily implies $\beta_a=l_{aa}=\pm 1$ and $k_{aa}=1$. On the other hand, if for a given $a$ we have $\beta_a=0$ then \eqref{eq:beta} implies
\begin{subequations}\label{eq:system2}
\begin{align}
x_{aa}=0,&\quad \beta_a=0\label{eq:system2a}\\
(4-\beta_b^2)x_{ab}=0,&\quad \beta_b\neq \beta_a=0.\label{eq:system2b}
\end{align}
\end{subequations}
Equations \eqref{eq:beta} and its derivations, \eqref{eq:beta2}$-$\eqref{eq:system2}, put constraints on the magnetic weight as well as on the form of $T_1$. 

Once we know the possible magnetic weights components, we turn to \eqref{eq:system1b}. It gives
$$
T_3=\frac{1}{8i}\sum_{abc}e_{ab}\otimes(\beta_aJx_{ac}x_{cb}-2\beta_cx_{ac}Jx_{cb}+\beta_bx_{ac}x_{cb}J).
$$
Comparing this last equation to \eqref{eq:T3}, multiplying on the right by $-J$ and using $JxJ=kx$ we arrive to the remaining fundamental constraints,
\begin{equation}\label{eq:beta4}
4\beta_a\delta_{ab}=\sum_c(\beta_ak_{ac}k_{cb}+2\beta_ck_{cb}+\beta_b)x_{ac}x_{cb},
\end{equation}
which determine the nontrivial blocks $x_{ab}$. The solutions of \eqref{eq:beta} and \eqref{eq:beta4} completely determine the generators $T_1$ and $T_3$.


\section{Classes of solutions for $SU(4)\rightarrow SO(4)$}\label{sec:classes}

In this section we find all possible magnetic weights as well as the explicit form of the $su(2)$ subalgebras associated to the monopoles. For simplicity we shall focus on the case $m=2$ i.e. $SU(4)\rightarrow SO(4)$. We show that there are three classes of solutions which we shall call fundamental, isoclinic and three-to-one embeddings.

\subsection{Fundamental embeddings}\label{subsec:fundamental}

We call fundamental embeddings those whose magnetic weight has only one nonvanishing component, namely $\beta_p$, and only $x_{pp}\neq 0$. The general conditions \eqref{eq:blocks} and $x_{ab}^\dagger=x_{ba}$ result in
\begin{equation}\label{eq:x11}
x_{pp}=\left(
\begin{array}{rr}
u&v\\
v&-u
\end{array}
\right),\quad u,v\in\mathbb R.
\end{equation}
and according to the discussion below \eqref{eq:beta3} we have $\beta_b=l_{pp}=\pm 1$ and $k_{pp}=1$. Equations \eqref{eq:beta4} yields
$$x_{pp}x_{pp}=1,$$ 
which gives $u^2+v^2=1$. Thus the general solution for the only nonvanishing block is
$$x_{pp}=\cos\chi\sigma_1+\sin\chi\sigma_3,$$
where $\chi$ is an arbitrary angle. This determines the general form of the fundamental embeddings. From \eqref{eq:T3}, \eqref{eq:T2} and \eqref{eq:T1} we obtain the $su(2)$ generators for a fundamental monopole as
\begin{equation}\label{eq:fundamental}
\begin{aligned}
T_1^{\beta_ae_a}&=\frac{1}{2}e_{aa}\otimes(\sin\chi\sigma_1+\cos\chi\sigma_3),\\
T_2^{\beta_ae_a}&=\frac{\beta_a}{2}e_{aa}\otimes(\cos\chi\sigma_1-\sin\chi\sigma_3),\\
T_3^{\beta_ae_a}&=\frac{\beta_a}{2}e_{aa}\otimes\sigma_2,
\end{aligned}
\end{equation}
with $\beta_a=\pm 1$ and $\beta=\beta_ae_a$. Notice that these equations hold for any $m$ and that the fundamental monopoles obtained in \cite{KL2010,KL2013} can be obtained from \eqref{eq:fundamental} by making $\cos\chi=0$ or $\cos\chi=\pi$ for $\beta_a=1$ or $\beta_a=-1$, respectively. Moreover, this angle is playing the same role as the free angle in \eqref{eq:tHP1}.

\subsection{Isoclinic embeddings}\label{subsec:isoclinic}

For the next case we consider that all magnetic weight components and all blocks $x_{ab}$ are nonzero. Then $\beta_a=l_{aa}=\pm 1$, $a=1,2$. There are two possibilities to be considered. The first one is when $\beta_1=l_{11}=\beta_2$. In this case, equations \eqref{eq:beta2} lead to $k_{ab}=1$. Plugging these quantities into \eqref{eq:beta4} we obtain three independent equations, namely
\begin{subequations}\label{eq:isoclicic}
\begin{align}
x_{11}x_{11}+x_{12}x_{21}&=1,\label{eq:isoclinica}\\
x_{21}x_{12}+x_{22}x_{22}&=1,\label{eq:isoclinicb}\\
x_{11}x_{12}+x_{12}x_{22}&=0.\label{eq:isoclinicc}
\end{align}
\end{subequations}
Since $x_{aa}$ has the same form as \eqref{eq:x11} we have 
$$x_{aa}x_{aa}=\rho_a^2,\quad\rho_a\in\mathbb{R}.$$
On the other hand,
$$x_{12}=\left(
\begin{array}{rr}
u_{12}&v_{12}\\
v_{12}&-u_{12}
\end{array}
\right).$$
Knowing these, we read from the diagonal elements of \eqref{eq:isoclinica} and \eqref{eq:isoclinicb} that
$$\rho_1^2=\rho_2^2=\cos^2\eta,\quad \rho_{12}^2\equiv |u_{12}|^2+|v_{12}|^2=\sin^2\eta,$$
for an angle $\eta$. In this way, 
$$x_{aa}=\cos\eta\left(
\begin{array}{rr}
\cos\chi_a&\sin\chi_a\\
\sin\chi_a&-\cos\chi_a
\end{array}
\right),$$
for new angles $\chi_1$ and $\chi_2$. From the off-diagonal elements we get $u_{12}v_{12}^\ast$ is real which means $u_{12}$ and $v_{12}$ have the same phase, namely $\psi$, and therefore
$$x_{12}=e^{i\psi}\sin\eta\left(
\begin{array}{rr}
\cos\gamma&\sin\gamma\\
\sin\gamma&-\cos\gamma
\end{array}
\right).$$
Plugging these blocks into \eqref{eq:isoclinicc} we check that the most general solution gives $\gamma=(\chi_1+\chi_2+\pi)/2$. This fixes the final form of the block $x_{12}$ and we are finally able to write the $su(2)$ generators as
\begin{equation}\label{eq:isoclinic1}
\begin{aligned}
T_1^{\pm(e_1+e_2)}&=\frac{1}{2}\begin{pmatrix}
\cos \eta (\sin \chi_1 \sigma_1+\cos \chi_1 \sigma_3)
&
e^{i\psi}\sin \eta (\cos\zeta\sigma_1-\sin\zeta\sigma_3)\\
e^{-i\psi}\sin \eta (\cos\zeta\sigma_1-\sin\zeta\sigma_3)
&
\cos \eta (\sin \chi_2 \sigma_1+\cos \chi_2 \sigma_3)
\end{pmatrix},\\
T_2^{\pm(e_1+e_2)}&=\pm\frac{1}{2}\begin{pmatrix}
\cos \eta (\cos \chi_1 \sigma_1 - \sin \chi_1 \sigma_3)
&
-e^{i\psi}\sin \eta (\sin\zeta\sigma_1+\cos\zeta\sigma_3)\\
-e^{-i\psi}\sin \eta (\sin\zeta\sigma_1+\cos\zeta\sigma_3)
&
\cos \eta (\cos \chi_2 \sigma_1 - \sin \chi_2 \sigma_3)
\end{pmatrix},\\
T_3^{\pm(e_1+e_2)}&=\pm\frac{1}{2}\begin{pmatrix}
\sigma_2& 0 \\
0& \sigma_2
\end{pmatrix},
\end{aligned}
\end{equation}
where $\zeta\equiv (\chi_1+\chi_2)/2$. It is straightforward to check that indeed these generators form an $su(2)$ algebra. A similar computation for $\beta_1=l_{11}=-\beta_2$ yields
\begin{equation}\label{eq:isoclinic2}
\begin{aligned}
T_1^{\pm(e_1-e_2)}&=\frac{1}{2}\begin{pmatrix}
\cos \eta ( \sin \chi_1 \sigma_1+\cos \chi_1 \sigma_3)
&
e^{i\psi}\sin \eta (\sin \xi \sigma_0-i\cos \xi \sigma_2)\\
e^{-i\psi}\sin \eta (\sin \xi \sigma_0+i\cos \xi \sigma_2)
&
\cos \eta (\sin \chi_2 \sigma_1+\cos \chi_2 \sigma_3 )
\end{pmatrix},\\
T_2^{\pm(e_1-e_2)}&=\pm\frac{1}{2}\begin{pmatrix}
\cos \eta (\cos \chi_1 \sigma_1 - \sin \chi_1 \sigma_3)
&
-e^{i\psi}\sin \eta (\cos \xi \sigma_0+i\sin \xi \sigma_2)\\
-e^{-i\psi}\sin \eta (\cos \xi \sigma_0-i\sin \xi \sigma_2)
&
\cos \eta (-\cos \chi_2 \sigma_1 + \sin \chi_2 \sigma_3)
\end{pmatrix},\\
T_3^{\pm(e_1-e_2)}&=\pm\frac{1}{2}\begin{pmatrix}
\sigma_2& 0 \\
0& -\sigma_2
\end{pmatrix},
\end{aligned}
\end{equation}
with $\xi=(\chi_2-\chi_1)/2$ and $\sigma_0$ the identity $2\times 2$ matrix. In both cases the requirement that neither $T_1$ nor $T_2$ annihilates the vacuum state implies $T_i^T\neq -T_i$, $i=1,\ 2$. This in turn excludes the possibilities where $\eta=\pi/2$ and $\psi=\pi/2,\ 3\pi/2$, for all $\chi_a$.

The form of $T_3$ in both cases motivates naming these embeddings as isoclinic embeddings. The $T_3$ in \eqref{eq:isoclinic1} generates a left-isoclinic rotation in four-dimensional space while $T_3$ in \eqref{eq:isoclinic2} generates a right-isoclinic rotation in the same space. The known $\mathbb Z_2$ monopoles constructed by \eqref{eq:KL2} can be recovered by setting $\eta=\chi_1=\chi_2=0$ and $\eta=0$, $\chi_1=\chi_2=\pi$ in \eqref{eq:isoclinic1}. On the other hand the  right-isoclinic embedding introduce a novelty since there is no solution in \cite{KL2010,KL2013} mixing fundamental monopoles associated to positive and negative magnetic weights.

\subsection{Three-to-one embeddings}\label{subsec:3-1}
There is one nontrivial possibility remaining which corresponds to all $x_{ab}\neq 0$ except for a single diagonal block. We will take it to be the second one for definiteness. Then $\beta_1=l_{11}$ and by \eqref{eq:beta3},
$$\beta_2=l_{21}(2-l_{12}l_{11}).$$
In order to avoid falling back to one of the previous families we must have $\beta_2\neq \pm 1$ which implies into $l_{12}=-l_{11}$ and $\beta_2= 3l_{21}=\pm 3$. Substituting these values in \eqref{eq:beta2} we find $k_{12}=-l_{11}l_{21}$. Now we are in place to find the nonvanishing blocks $x_{ab}$. We do so by solving \eqref{eq:beta4}, whose the two nontrivial equations form the system
\begin{align*}
1&=x_{11}x_{11}-x_{12}x_{21},\\
3&=x_{21}x_{12}.
\end{align*}
By inserting the block
$$
x_{12}=\left(
\begin{array}{cc}
u_{12}&v_{12}\\
k_{12}v_{12}&-k_{12}u_{12}
\end{array}
\right),
$$
into the second equation of the system it is solved for
$$
x_{12}=\sqrt 3e^{i\psi}\left(
\begin{array}{cc}
\cos\chi_2&\sin\chi_2\\
k_{12}\sin\chi_2&-k_{12}\cos\chi_2
\end{array}
\right).
$$
The first equation of the system becomes $x_{11}x_{11}=4$ and therefore
$$x_{11}=2(\sin\chi_1\sigma_1+\cos\chi_1\sigma_3).$$
Finally the $su(2)$ subalgebras are
\begin{align*}
T_1^{\pm(e_1+3e_2)}&=\frac{1}{2}\begin{pmatrix}
2(\sin\chi_1\sigma_1+\cos\chi_1\sigma_3)&
\!\!\!\!\!\!\!\!\sqrt{3}e^{i\psi}(\cos\chi_2\sigma_0+i\sin\chi_2\sigma_2) \\
\sqrt{3}e^{-i\psi}(\cos\chi_2\sigma_0-i\sin\chi_2\sigma_2) 
& 0 \end{pmatrix},\\
T_2^{\pm(e_1+3e_2)}&=\pm\frac{1}{2}\begin{pmatrix}
2(\cos\chi_1\sigma_1-\sin\chi_1\sigma_3)&
\!\!\!\!\!\!\!\!\sqrt{3}e^{i\psi}(i\cos\chi_2\sigma_2-\sin\chi_2\sigma_0) \\
\sqrt{3}e^{-i\psi}(-i\cos\chi_2\sigma_2-\sin\chi_2\sigma_0)
& 0 \end{pmatrix},\\
T_3^{\pm(e_1+3e_2)}&=\pm\frac{1}{2}\begin{pmatrix}
\sigma_2 & 0\\
0 & 3\sigma_2
\end{pmatrix}.
\end{align*}
and
\begin{align*}
T_1^{\pm(e_1-3e_2)}&=\frac{1}{2}\begin{pmatrix}
2(\sin\chi_1\sigma_1+\cos\chi_1\sigma_3)&
\!\!\!\!\!\!\!\!\sqrt{3}e^{i\psi}(\sin\chi_2\sigma_1+\cos\chi_2\sigma_3) \\
\sqrt{3}e^{-i\psi}(\sin\chi_2\sigma_1+\cos\chi_2\sigma_3) 
& 0 \end{pmatrix},\\
T_2^{\pm(e_1-3e_2)}&=\pm\frac{1}{2}\begin{pmatrix}
2(\cos\chi_1\sigma_1-\sin\chi_1\sigma_3)&
\!\!\!\!\!\!\!\!\sqrt{3}e^{i\psi}(\cos\chi_2\sigma_1-\sin\chi_2\sigma_3) \\
\sqrt{3}e^{-i\psi}(\cos\chi_2\sigma_1-\sin\chi_2\sigma_3) 
& 0 \end{pmatrix},\\
T_3^{\pm(e_1-3e_2)}&=\pm\frac{1}{2}\begin{pmatrix}
\sigma_2 & 0\\
0 & -3\sigma_2
\end{pmatrix}.
\end{align*}
for $\beta=\pm (1,3)$ and $\beta=\pm (1,-3)$, respectively. We shall refer to these two embeddings as three-to-one because $T_3$ generates a simultaneous rotation of two orthogonal planes, the second plane rotating three times faster than the first one. Analogous results hold for $\beta=\pm (3,1)$ and $\beta=\pm (3,-1)$. The associated monopoles in any of these embeddings are not predicted in \cite{KL2010,KL2013}.
\\ \\
For the sake of completeness we rule out the case $x_{11}=x_{22}=0$ and $\beta_p=0$ for a given $p$. If, for instance, $\beta_2=0$, \eqref{eq:beta3} gives $\beta_1=\pm 2$ and by \eqref{eq:system1b}
$$T_3=e_{11}\otimes[x_{12}x_{12}^\dagger,\sigma_2],$$
which is a contradiction since the rhs is not Hermitian. Thus there are no such embeddings. 

The asymptotic form of the obtained $\mathbb Z_2$ monopoles constructed with the obtained $su(2)$ subalgebras are given by \eqref{eq:general asymptotic fields}. The fields with unit winding number show the spherically symmetric hedgehog form
\begin{align*}
\phi(\theta,\varphi)&=v\frac{x_aT_a}{r},\\
W_i(\theta,\varphi)&=\epsilon_{aij}\frac{x_jT_a}{er^2},
\end{align*}
with $a=1,\, 2,\, 3$, as expected. It dos not show explicit dependence on the found continuous parameters. The BPS mass of these monopoles can be found in the same way it is done in \cite{KL2013}. Once again it will not depend on these parameters. The eventual relation of these parameters with zero modes is a subject of further research.

\section{Conclusions}\label{sec:conclusions}

In this work we investigate possible $\mathbb Z_2$ monopoles in $su(2)$ embeddings of larger algebras. The method developed lead the main equations \eqref{eq:beta} and \eqref{eq:beta4}. These are general enough so as to classify all $su(2)$ subalgebras of $su(2m)$ bearing only one generator of the given $so(2m)$. In our case the unbroken algebra is invariant under Cartan automorphism and annihilates the vacuum configuration given by \eqref{eq:vacuum}. For $SU(4)\rightarrow SO(4)$ all nontrivial possibilities has been explicitly found and they correspond to the so-called fundamental, isoclinic and three-to-one embeddings. The devised method can be used for the most general case $SU(2m)\rightarrow SO(2m)$ and there are some straightforward generalizations like the fundamental embeddings, with magnetic weights $\beta=\pm e_a$, $a=1,\ldots,m$ and the isoclinic embeddings, $\beta=\sum_a n_ae_a$, $n=-1,\, 0,\, 1$, where two or more $n_i$ are nonzero.

\section*{Acknowledgments}

P. J. L. wishes to thank M. A. C. Kneipp for useful discussions and FAPESC for financial support. E. E. Q. is grateful to CNPq for financial support.

\bibliographystyle{ieeetr}
\bibliography{bibliografia}

\end{document}